# Nuclear Medicine AI in Action: The Bethesda Report (AI Summit 2024)

Arman Rahmim*; Departments of Radiology & Physics, University of British Columbia, Vancouver, Canada

Tyler J. Bradshaw; Department of Radiology, University of Wisconsin, Madison, USA

Guido Davidzon; Department of Radiology, Stanford University, Palo Alto, USA

Joyita Dutta; Department of Biomedical Engineering, University of Massachusetts Amherst, USA

Georges El Fakhri; Departments of Radiology & Biomedical Engineering and Bioinformatics & Data Science, Yale University, New Haven, USA

Munir Ghesani; United Theranostics, Princeton, USA

Nicolas A. Karakatsanis; Department of Radiology, Weill Cornell Medical College, New York, USA

Quanzheng Li; Department of Radiology, Massachusetts General Hospital, Harvard Medical School, Boston, USA

Chi Liu; Department of Radiology & Biomedical Imaging, Yale University, New Haven, USA

Emilie Roncali; Departments of Biomedical Engineering & Radiology, University of California, Davis, USA

Babak Saboury; Institute of Nuclear Medicine, Bethesda, USA

Tahir Yusufaly; Department of Radiology & Radiological Sciences, Johns Hopkins University, Baltimore, USA

Abhinav K. Jha; Department of Biomedical Engineering and Mallinckrodt Institute of Radiology, Washington University, St. Louis, USA

* Corresponding Author: Arman Rahmim, PhD; 675 W 10th Ave., Vancouver, BC, Canada; 604 404 7922
675 West 10th Ave; Office 6-112 Vancouver, BC, V5Z 1L3
Phone: 604-675-8262
arman.rahmim@ubc.ca

## Abstract

The 2nd Society of Nuclear Medicine & Molecular Imaging (SNMMI) AI Summit, organized by the SNMMI AI Task Force, took place in Bethesda, MD, on February 29-March 1, 2024. Bringing together various community members and stakeholders, and following up on a prior successful 2022 Summit, the theme was “AI in Action”. Key items discussed included emerging needs and tools for computational nuclear oncology, new frontiers in large language and generative models, defining the value proposition for the use of AI in nuclear medicine, open science including efforts for data and model repositories, and issues of reimbursement and funding. The primary efforts, findings, challenges, and next steps are summarized in this manuscript.

## Introduction

In the 2nd SNMMI AI Summit, >100 community members and stakeholders from academia, healthcare, industry, Food and Drug Administration (FDA), and National Institutes of Health gathered to discuss the emergence of AI in nuclear medicine. Building on the success of the 1st AI Summit, the summit featured two plenaries, panel discussions, talks from leading experts, and a roundtable discussion on key findings and next steps. The 2024 theme was “AI in Action”, with the goal of identifying opportunities and barriers to clinical implementation of AI. This report summarizes five topics of discussion, including visions and challenges to overcome, and does not represent the official views of the SNMMI.

## II. Computational nuclear oncology

We define computational nuclear oncology (CNO) as the application of computational methods, including AI, to enable improved assessment of absorbed doses in the present time, and to predict doses (*1*) as well as biological and clinical effects in future radiopharmaceutical therapy (RPT) cycles (*2*). Predictive modeling in RPT is challenging, since the processes underlying radiopharmaceutical delivery and efficacy span a broad range of length and time scales that happen beyond the current spatiotemporal resolution of diagnostic nuclear medicine imaging. Consequently, it is necessary to augment clinical images with additional computational modeling to account for these unobservable processes.

The CNO pipeline starts with pharmacokinetic models, parameterized using diagnostic imaging of a surrogate radiotracer, to predict biodistribution of the therapeutic agent. This information is coupled with particle transport codes to estimate energy transfer and cytotoxicity to biological targets. These outputs are fed into simulations of multicellular response, including normal organ wound healing and tumor eco-evolution (*3*), to interface with observable clinical endpoints. We anticipate a theranostic digital twin paradigm, with virtual representations of patients that are updated with real-world data for optimized personalized therapies, in line with ongoing developments in the larger field of image-guided digital twins for clinical oncology (*4*).

Major challenges include integration of existing tools, accounting for interactions and interdependencies of different processes, and empirical benchmarking to ensure robustness and interoperability. Given the labor-intensiveness of the upcoming challenges, automated approaches based on AI will be useful, if not necessary.

### III. Large language and generative models

A significant recent trend is the development of foundation models (*5*). In healthcare, large language models (LLMs), foundation models that can process or generate text, have shown particular promise. LLMs have captivated the general populace since the introduction of ChatGPT, with several versions released since, including GPT-5 (OpenAI), PaLM and Gemini (Google), and the Llama family (Meta).

LLMs can handle, mine, and generate text, including electronic health records (EHRs), clinical notes, and publications, thus automating the retrieval, extraction and curation of information from medical literature. Relatedly, recent efforts have leveraged natural language processing (NLP) to improve search quality in the PubMed database. Specifically, LitCovid emerged as a prominent NLP-powered resource (*6*). Other prominent NLP-powered tools include LitVar, for search and retrieval of genetic-variant-specific information from research articles, and PubTator, providing computer-annotated biomedical concepts, such as genes and mutations. Applications of LLMs to nuclear medicine are likely to mirror those in radiology, but further study is needed to appreciate any promises and challenges specific to nuclear medicine.

Generative AI models have shown promise in medical imaging, particularly with advances in diffusion models that have the ability to generate minority samples from unbalanced data distributions (*7*). In medicine, these models are being applied to synthetic data generation, image denoising, and image restoration. It is important to pay close attention to any assumptions built-in to these models before applying them in nuclear medicine. For example, nuclear-medicine images have signal-dependent noise distributions, in contrast to the signal-independent distribution often assumed. Inappropriate assumptions may cause generative models to hallucinate, introducing biases in the generated data.

### IV. Value propositions for the use of AI

Clinical adoption of AI will require that algorithms provide a higher clinical "value" over standard of care. The increased value may arise from several factors, such as increased accuracy and precision, reduction in time and money, or automation of clinically burdensome tasks.

Quantifying this value is challenging and requires thoughtful consideration, with rigorous validation on clinical tasks being crucial. An AI denoising technique may improve visual appeal of an image, but provide no additional clinical utility (e.g. in myocardial perfusion SPECT (*8*)). Similarly, evaluation of AI oncological PET segmentation procedure using task-agnostic metrics (e.g. Dice scores) may provide different interpretations than on clinically relevant tasks such as estimating metabolic tumor volume (*9*). Task-based evaluation is thus essential, a requirement extending to more general tasks, such as classification (*10*). A framework for such task-based evaluation has been outlined (*11*). Further clinical evaluation of algorithms may also be necessary. For example, there may be cases where segmentation algorithms detect a small foci of physiologic uptake that leads to false positives in patients without disease or in complete remission. Best practices to evaluate AI algorithms (RELAINCE guidelines) have been proposed by the AI Task Force (*12*), and the Journal of Nuclear Medicine now recommends papers on AI evaluation provide a quantitative claim of performance.

AI can assist physician interpretation of PET/CT, PET/MR and SPECT/CT scans. The recent surge in demand for clinical imaging is juxtaposed against a relatively stable supply of Board-Certified Professionals capable of interpreting scans (*13*). The latter are often mired in repetitive manual tasks instead of intellectual endeavors. Recent endeavors have explored the potential of AI towards addressing these challenges, such as through tumor segmentation and quantification of total metabolic tumor volume from [$^{18}$F]FDG-PET scans (*14*), or via FDA-approved software for automated delineating and quantification of abnormal prostate-specific membrane antigen (PSMA)-avid foci (*15*). More complex metrics like tumor volume in PSMA PET are increasingly recommended, e.g. by the new PROMISE framework, and AI can assist with these computations.

The amalgamation of images with other clinical digital data (e.g., pathology, laboratory results, clinical parameters, demographics, genomics, and regulatory guidelines) offers an opportunity for AI in report generation. In addition to highlighting abnormal findings, such reports could incorporate personalized therapeutic recommendations based on extracted patient-specific characteristics (Figure 1). For instance, PET/CT reports could incorporate personalized risk scores derived from integrative radiomic and clinical models. Such comprehensive staging approaches can help identify patients who may benefit from more aggressive initial treatments (*16*)(*17*).

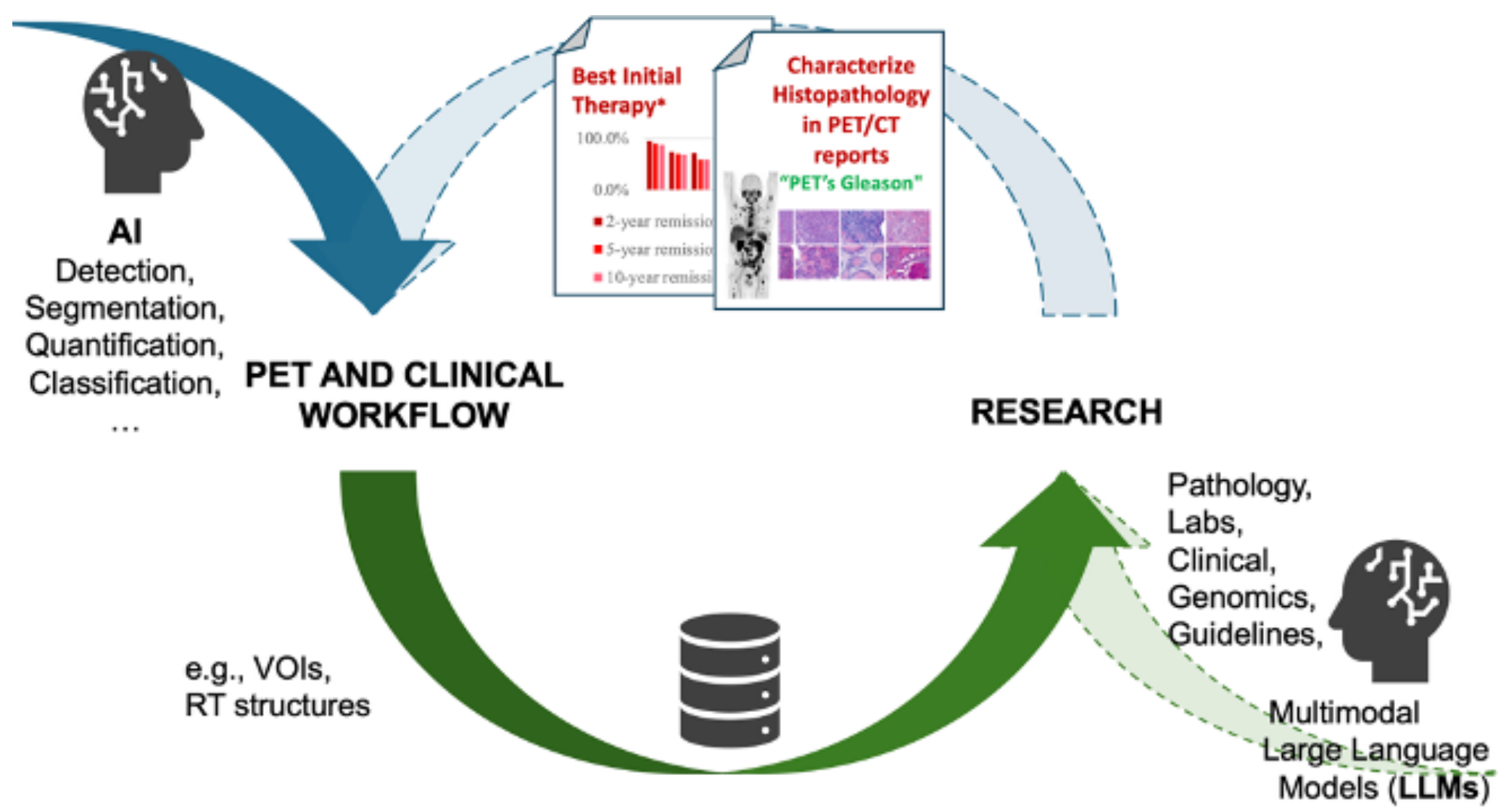

**Figure 1—Future PET/CT clinical reports, based on empiric results generated from multimodal AI models, can predict clinical outcomes and make recommendations for optimal therapies.**

An important but often neglected topic is how AI can help nuclear-medicine technologists, who manage patient workflows, interface with patients, acquire images, and conduct quality control. Here, AI could help in multiple ways, such as scheduling patient workflows, automating manual SPECT-CT alignment, and reducing radiation dose and/or scan time. As was insightfully pointed out at the summit, "AI will not replace nuclear medicine technologists but nuclear medicine technologists who know AI will replace those who do not."

AI may also reduce patient-care costs. For nuclear medicine in particular, CT-less attenuation compensation in cardiac PET and SPECT, where the CT scans are acquired only for attenuation compensation, and reducing radiation dose and acquisition time in PET and SPECT scans. In addition to saving costs related to imaging studies, about one third of net saving opportunities achievable by AI can be focused on administrative costs (*18*). There is an important need for rigorous investigation on cost-effectiveness of AI technologies through prospective studies.

## V. Open science

The growth in AI publications in nuclear medicine has not been accompanied by an increase in shared models. This gap prevents independent validation, necessary for realizing the full clinical potential of these models. Encouraging code-sharing practices will be essential for promoting trustworthy, reproducible AI and driving wider clinical adoption. Fostering a culture of shareability can also support the use of restricted-access archived data for federated training and external validation of publicly available AI models.

To this end, the AI Task Force has undertaken a systematic online literature review, and identified shareable AI models with publicly available executable code, source code, and documentation. These models have been organized into an open-access web database on the NMMItools.org platform, featuring structured templates, classifications (e.g., task, disease, model type), and tagged keywords for easy navigation. Automated literature searches and an interactive feedback tool ensure continuous updates and collaboration between developers and users. This open, interactive platform aims to build a thriving community, promote awareness of shareable AI models, and encourage responsible code sharing practices.

## VI. Reimbursement and funding

Responsible adoption of healthcare AI requires that AI systems that benefit patients and populations are incentivized financially at a consistent and sustainable level. The decision by providers and health care systems on whether to adopt and deploy a specific healthcare AI is greatly influenced by the factors of reimbursement, coverage, and medical liability.

Reimbursement for AI algorithms needs to take into account extra expenditures if an AI service is not provided, compared to when provided. A late diagnosis increases treatment costs and a poorer outcome has adverse financial consequences. The cost-benefit threshold for the value of AI is derived from what payers are currently paying for the service provided by a human provider, such as a specialist. Many of these services are valued based on the provider's expertise, time spent, and ancillary services required. It is derived from the current total value for a subset of the population assigned to the entire population getting the service, at the population level.

Technologies grouped under the umbrella of Software as a Service (SaaS) rely on complex algorithms or statistical predictive modeling to aid in the diagnosis or treatment of a patient's condition, and which the providers pay for either on a subscription or per-use basis. One such example is fractional flow reserve (FFR) Derived from Computed Tomography, entailing proprietary data analysis to generate a 3D image of a patient's coronary arteries, identify FFR, and determine the need for invasive coronary angiography. There is considerable promise of SaaS in nuclear medicine.

A few years ago, Centers for Medicare & Medicaid Services released a proposed rule for the Medicare Coverage of Innovative Technology (MCIT), a pathway to providing Medicare payment for technology that the FDA has deemed a breakthrough device. Such devices must provide more effective treatment or diagnosis of a life-threatening or irreversibly debilitating condition, beyond that provided by any other other cleared or approved alternatives. Unfortunately, it will be difficult to develop nuclear medicine AI algorithms that would meet MCIT requirements.

While radiology dominates most FDA-approved AI algorithms, less than 10% are applicable in diagnostic and therapeutic nuclear medicine, molecular imaging and RPT. Thus, the nuclear medicine community needs to work collaboratively to introduce new algorithms and demonstrate their value.

## Conclusions and Call to Action

The SNMMI 2024 AI Summit emphasized the importance of demonstrating AI's "value," crucial for establishing reimbursement models. Achieving this requires collaboration among nuclear medicine stakeholders—physicians, physicists, technologists, and computational scientists—and research on AI's role in reducing patient care costs and supporting global healthcare. Highlighting AI's potential to address clinically significant challenges, such as in computational nuclear oncology, could showcase its impact on treatment and drive further research.

Another takeaway was addressing the challenge of data availability in nuclear medicine. Data is needed to train and then rigorously validate AI algorithms. Further, to establish generalizability, testing on multi-center data is important. To achieve these goals, there is a strong need for centers to join hands to share data through approaches such as centralized data repositories. However, it was also recognized that there are challenges with sharing data, and approaches such as federated learning may be considered.

The summit highlighted the need for an educational program on AI within the nuclear medicine community, targeting physicians, physicists, computational scientists, technologists, and industry partners. This program would offer multidisciplinary training for residents, students, postdocs, and engineers, fostering collaboration between AI developers and clinicians to align AI solutions with clinical needs. There was also interest in developing an SNMMI AI certificate program.

Additionally, rigorous validation of AI algorithms was emphasized, with the proposed RELAINCE guidelines that can serve as a foundation for validating AI on clinical tasks.

**Acknowledgements**
We give special thanks to Bonnie Clarke and Lance Burrell with significant efforts to organize SNMMI AI task force meetings as well as the 2024 AI Summit.

**Disclosure Statement**
No relevant conflicts of interest.

**KEY POINTS**

**QUESTION:**

What are the challenges and opportunities, identified at the 2024 SNMMI AI Summit, towards clinical implementation of artificial intelligence (AI) in nuclear medicine?

**PERTINENT FINDINGS:**
At the AI Summit, many experts examined AI applications in nuclear medicine, including computational nuclear oncology tools, large language and generative models, clinical value assessment, open science considerations, and issues of reimbursement. Key insights included the need for validated, clinically relevant AI tools that demonstrate “value”, integration of multiscale modeling for radiopharmaceutical therapy, development of open-access repositories to improve reproducibility, and need for significantly improved collaborations, data sharing and educational efforts.

**IMPLICATIONS FOR PATIENT CARE:**
AI has the potential to improve nuclear medicine through enhanced precision, efficiency, and personalization, but achieving clinical impact depends on rigorous validation, data sharing, and supportive reimbursement frameworks.